\documentclass{iopart}
\usepackage{iopams,citesort,graphicx,tabularx}  
\begin{document}

\newcommand{\rd}{\mathrm{d}}
\newcommand{\expct}[1]{\langle #1 \rangle}
\newcommand{\Expct}[1]{\left\langle #1 \right\rangle}
\newcommand{\cum}[1]{{\langle #1 \rangle}_{\rm c}}
\newcommand{\Cum}[1]{{\left\langle #1 \right\rangle}_{\rm c}}
\newcommand{\diff}[2]{\frac{\mathrm{d} #1}{\mathrm{d} #2}}
\newcommand{\prt}[2]{\frac{\partial #1}{\partial #2}}
\renewcommand{\(}{\left(}
\renewcommand{\)}{\right)}
\renewcommand{\[}{\left[}
\renewcommand{\]}{\right]}
\newcommand{\im}{\mathrm{Im}}
\newcommand{\re}{\mathrm{Re}}
\newcommand{\Std}{\mathrm{Std}}
\newcommand{\cdf}{\mathrm{cdf}}
\newcommand{\unit}[1]{~\mathrm{#1}}

\title[Circular interfaces in an off-lattice Eden model]{Statistics of circular interface fluctuations\\ in an off-lattice Eden model}

\author{Kazumasa A Takeuchi}

\address{Department of Physics, the University of Tokyo,
 7-3-1 Hongo, Bunkyo-ku, Tokyo 113-0033, Japan.}
\ead{kat@kaztake.org}
\begin{abstract}
Scale-invariant fluctuations of growing interfaces are studied
 for circular clusters of an off-lattice variant of the Eden model,
 which belongs to the $(1+1)$-dimensional
 Kardar-Parisi-Zhang (KPZ) universality class.
Statistical properties of the height (radius) fluctuations
 are numerically determined
 and compared with the recent theoretical developments
 as well as the author's experimental result
 on growing interfaces in turbulent liquid crystal
 [K. A. Takeuchi and M. Sano, arXiv:1203.2530].
We focus in particular on analytically unsolved properties
 such as the temporal correlation function
 and the persistence probability in space and time.
Good agreement with the experiment is found
 in characteristic quantities for them,
 which implies that the geometry-dependent universality 
 of the KPZ class holds here as well,
 but otherwise a few dissimilarities are also found.
Finite-time corrections in the cumulants of the distribution
 are also studied and
 shown to decay, for the mean, as $t^{-2/3}$
 within the time window of the simulations,
 instead of $t^{-1/3}$
 which arises as the typical leading term in the previously known cases.
\end{abstract}



\section{Introduction}

Growing interfaces driven by local stochastic interactions
 constitute a prototypical example where
 scale invariance leads to universality
 even in systems far from equilibrium \cite{Barabasi.Stanley-Book1995}.
In their seminal work \cite{Kardar.etal-PRL1986},
 Kardar, Parisi, and Zhang (KPZ) proposed
 a continuum equation for describing such growing interfaces, which reads
\begin{equation}
 \prt{}{t}h(x,t) = \nu \nabla^2 h + \frac{\lambda}{2}(\nabla h)^2 + \xi(x,t)  \label{eq:KPZEqDef}
\end{equation}
 with the local height $h(x,t)$ of the interfaces and
 the white Gaussian noise $\expct{\xi(x,t)} = 0$
 and $\expct{\xi(x,t)\xi(x',t')} = D\delta(x-x')\delta(t-t')$.
They derived a set of characteristic exponents
 $\alpha = 1/2$, $\beta = 1/3$ and $z \equiv \alpha/\beta = 3/2$
 that describe scale-invariant fluctuations of the interfaces
 in $1+1$ dimensions,
 in agreement with the earlier theoretical studies on related problems
 \cite{Forster.etal-PRA1977,vanBeijeren.etal-PRL1985}.
These exponents are universal
 as widely confirmed by theoretical and numerical work
 \cite{Barabasi.Stanley-Book1995} as well as by a few experiments
 (see \cite{Takeuchi.Sano-PRL2010,Takeuchi.etal-SR2011,Takeuchi.Sano-a2012}
 and references therein),
 constituting the KPZ universality class.
Recently, studies in this context entered an unprecedented phase,
 when Johansson rigorously derived the asymptotic distribution function
 for a model in the KPZ class \cite{Johansson-CMP2000}
 on tha basis of related combinatorial problems \cite{Baik.etal-JAMS1999};
 this spearheaded remarkable theoretical achievements
 marked in the last decade
 (for recent reviews,
 see \cite{Kriecherbauer.Krug-JPA2010,Sasamoto.Spohn-JSM2010,Corwin-RMTA2012}).
Their main conclusions are twofold:
 (i) The distribution function and the spatial correlation function
 were obtained analytically for a few solvable models,
 including the KPZ equation \cite{Sasamoto.Spohn-PRL2010,Sasamoto.Spohn-NPB2010,Amir.etal-CPAM2011,Calabrese.etal-EL2010,Dotsenko-EL2010,Calabrese.LeDoussal-PRL2011,Imamura.Sasamoto-a2011},
 and revealed nontrivial connection to random matrix theory.
 (ii) These are expected to be universal, but nevertheless
 depend on the global shape of the interfaces,
 e.g., whether they are curved or flat,
 as first pointed out by Pr\"ahofer and Spohn \cite{Prahofer.Spohn-PRL2000}.
In particular, the asymptotic distribution function
 for the curved interfaces is given
 by the largest-eigenvalue distribution of large random matrices
 in Gaussian unitary ensemble (GUE)
 \cite{Johansson-CMP2000,Prahofer.Spohn-PRL2000},
 called the GUE Tracy-Widom distribution \cite{Tracy.Widom-CMP1994},
 and for the flat interfaces by the corresponding distribution
 for Gaussian orthogonal ensemble (GOE) \cite{Prahofer.Spohn-PRL2000}.

This detailed yet geometry-dependent universality of the KPZ class
 was recently underpinned by an experiment on growing interfaces
 of topological-defect turbulence in nematic liquid crystal,
 in which the author and his coworker identified
 the analytically obtained distribution functions
 and the spatial correlation functions for both circular and flat interfaces
 \cite{Takeuchi.Sano-PRL2010,Takeuchi.etal-SR2011,Takeuchi.Sano-a2012}.
In \cite{Takeuchi.Sano-a2012} they further advanced their analyses
 to study statistical properties
 that remain analytically unsolved even for the solvable models,
 in particular those related to
 the temporal correlation of the interface fluctuations.
This revealed that the difference between the circular and flat interfaces
 is not restricted to the distribution and spatial correlation functions,
 but actually arises in many other quantities as well,
 sometimes even with qualitative differences,
 as summarized in table 1 of \cite{Takeuchi.Sano-a2012}.
These quantities --- specifically,
 the temporal correlation function and
 the temporal and spatial persistence probabilities ---
 are expected to be universal in the asymptotic limit
 when appropriately rescaled.

Here, as an approach complementary to the experiment,
 the author performs the same analyses for a numerical model,
 namely an off-lattice variant of the Eden model
 \cite{Eden-Proceedings1961,Barabasi.Stanley-Book1995}
 introduced in the present paper,
 and provides a numerical case study on the universality
 in those unsolved quantities.
Focus in the present study is set on the circular interfaces,
 which are numerically much less studied than the flat case
 and are also more delicate:
 because the circumference, or the system size of the interface,
 is by construction finite at finite times and grows with time,
 the ensemble and the spatial averages lead to different results
 when they are used to define the statistical quantities
 \cite{Takeuchi.Sano-a2012}.
Using the spatial average instead of the ensemble one yields
 false estimates even for the growth exponent $\beta$
 \cite{Takeuchi.Sano-a2012}. 
Since this point has not been explicitly recognized in past studies,
 it is important to provide a set of reliable numerical estimates
 for the circular interfaces of a numerical model ---
 this is the aim of the present paper.

\section{Model}

Only few numerical models are known to produce circular interfaces,
 most of which are variants of the Eden model.
The Eden model was originally defined on a lattice.
Starting with a seed particle at the origin,
 at each time step one adds a new particle on a randomly chosen
 site on the perimeter of the cluster
 \cite{Eden-Proceedings1961,Barabasi.Stanley-Book1995}.
This however turned out to induce anisotropy due to the lattice structure
 \cite{Barabasi.Stanley-Book1995,Freche.etal-JPA1985}
 unless a specific growth rule is introduced,
 such as the growth probability dependent
 on the number of the occupied nearest-neighbors
 \cite{Paiva.FerreiraJr-JPA2007,Alves.etal-EL2011}.
Off-lattice versions of the Eden model have therefore been
 considered occasionally in the literature
 \cite{Wang.etal-JPA1995,FerreiraJr.Alves-JSM2006,Kuennen.Wang-JSM2008,Alves.etal-EL2011},
 but in most cases time $t$ is measured by the global radius $\expct{h}$
 of the radial cluster, which is not correct at finite times
 because the two quantities are in general connected
 by $\rd\expct{h}/\rd t \simeq v_\infty + c_vt^{-2/3}$
 \cite{Krug.etal-PRA1992,Takeuchi.Sano-PRL2010,Takeuchi.etal-SR2011,Takeuchi.Sano-a2012}.
To resolve these problems, as well as to study a numerical model
 that the author considers better corresponds to a coarse-grained picture
 for the experimentally investigated growth
 of the liquid-crystal turbulence
 \cite{Takeuchi.Sano-PRL2010,Takeuchi.etal-SR2011,Takeuchi.Sano-a2012},
 a new variant of the off-lattice Eden model is introduced and studied
 in the present paper.

\begin{figure}[t]
 \begin{center}
  \includegraphics[clip]{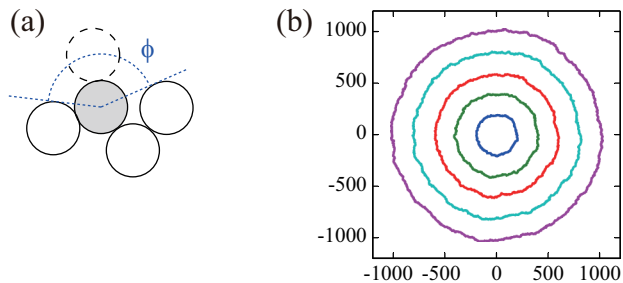}
  \caption{Off-lattice Eden model. (a) Sketch of the model. A new particle (dashed) is being added to its ancestor (shaded). In Alves \textit{et al.}'s model \cite{Alves.etal-EL2011,FerreiraJr.Alves-JSM2006,Alves.etal-BJP2008}, a random number for the angular position of the new particle is generated within $[\pi/6,\phi-\pi/6]$, where $\phi$ is defined in the sketch and $\pi/6$ is due to the finite radius of the new particle. In contrast, in our model, the random number is always drawn from $[0,2\pi)$ and then the attempt of generating the new particle is accepted with probability $(\phi-\pi/3)/2\pi$. (b) Typical cluster produced by our Eden model. The active particles, or the interface, at $t = 405, 803, 1174, 1592, 1999$ are shown.}
  \label{fig:Model}
 \end{center}
\end{figure}

The model is defined as follows.
First, place a round particle of unit diameter
 at the origin of two-dimensional continuous space.
Identical particles are then added one by one according to the following steps,
 where $N_{\rm a}$ is the number of the active particles
 as defined below.
The initial particle is active, hence $N_{\rm a} = 1$ at $t=0$.
(1) Choose one of the $N_{\rm a}$ active particles randomly.
(2) Attempt to place a new particle on its border,
 in a direction randomly chosen in the range $[0,2\pi)$.
(3) Place this particle if it does not overlap any other particles,
 otherwise abandon the attempt [figure \ref{fig:Model}(a)].
(4) Label as inactive those particles
 to which no particle can be added any more for lack of empty space,
 thereby excluding them from the list of the $N_{\rm a}$ active particles.
Since we are interested in the dynamics of the interface,
 we also exclude the particles left in the bulk,
 which are surrounded by an outer closed loop of adjacent particles.
Here, particles are regarded as adjacent if the distance between them
 is shorter than $\sqrt{3}$ because in this case an inner particle
 cannot influence outer ones in any manner.
Note also that the inactive particles remain obstacles
 for the active ones; in practice,
 this is realized by recording the angles of the inactive neighborhood
 for each active particle.
(5) Increase time $t$ by $1/N_{\rm a}$,
 whether the new particle is placed or not.
This choice indicates that
 each particle has a chance to add a neighbor
 once per unit time on average.
Therefore, it is statistically equivalent to use
 the whole particle number $N$ instead of $N_{\rm a}$
 for all the steps above,
 but the use of the active particle number $N_{\rm a}$
 speeds up the simulations significantly.

This model resembles the off-lattice Eden model studied by
 Alves \textit{et al.} \cite{Alves.etal-EL2011}
 (also described in \cite{FerreiraJr.Alves-JSM2006,Alves.etal-BJP2008})
 except for the criterion of the inner particles
 and for the way a new particle is added to the randomly chosen ancestor.
On the former point, Alves \textit{et al.} inactivated
 the particles inside a central core of radius $0.8\bar{r}$
 \cite{FerreiraJr.Alves-JSM2006,Alves.etal-BJP2008},
 where $\bar{r}$ is the mean distance of the particles on the interface
 from the origin, whereas in our model it is determined
 by the purely geometrical consideration,
 for which there is no risk of false inactivation.
The latter point concerns the step (2) in the above procedure.
When determining the direction to place a new particle,
 Alves \textit{et al.} drew a random number
 from such angles that correspond to the empty neighborhood
 of the ancestor, while we always choose an angle within $[0,2\pi)$
 and then judge whether the particle can be placed or not
 [figure \ref{fig:Model}(a)].
This guarantees local isotropy of the growth process.
The author also considers that, in coarse-grained scales, it better corresponds
 to the growth of topological defects in the liquid-crystal turbulence
 \cite{Takeuchi.Sano-PRL2010,Takeuchi.etal-SR2011,Takeuchi.Sano-a2012}.

The data presented below are obtained from 3000 independent simulations
 of the above-defined Eden model up to $t=1999$,
 which roughly corresponds
 to the final cluster size of $N = 2 \times 10^6$
 if one does not discard the inner particles.
Figure \ref{fig:Model}(b) shows a typical growing cluster;
 it displays the active particles at different times,
 labeled by $i$ below,
 which form the interface at each time by construction.
The local interface height $h(x,t)$ is then determined
 from the distance $r_i$ between the origin and the active particle $i$
 at a given angular position.
More specifically, the full azimuthal range $[0,2\pi)$
 is divided into $\lfloor 2\pi\expct{r_i} \rfloor$ bins,
 where $\expct{r_i}$ is the ensemble average of $r_i$,
 or the mean radius of the clusters at the given time,
 and $\lfloor\cdots\rfloor$ represents the integer part;
 then, for each bin and cluster, a single value of $h(x,t)$ is assigned
 as the mean of $r_i$ within the corresponding range of angular positions.
Note that the width of the single bin
 is then roughly equal to the diameter of the particles.
The lateral coordinate $x$ is defined
 along the mean shape of the clusters,
 i.e., the circle of circumference $2\pi\expct{r_i}$.

\section{Results}

\begin{figure}[t]
 \begin{center}
  \includegraphics[clip]{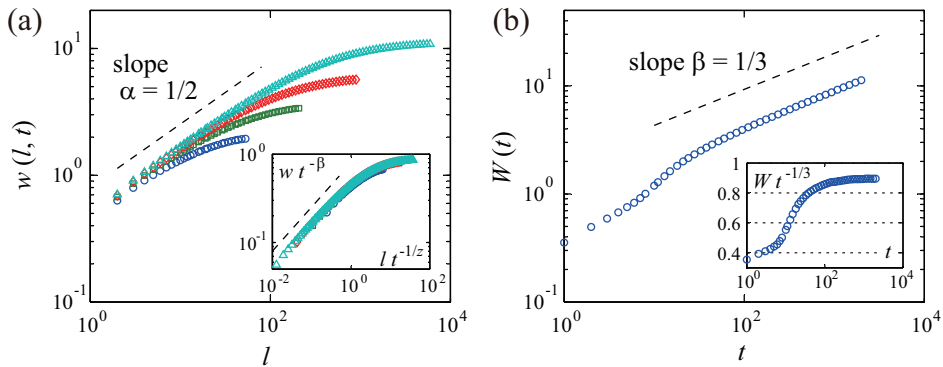}
  \caption{Family-Vicsek scaling. (a) Width $w(l,t)$ against length $l$ at different times $t=24, 82, 322, 1999$ (from bottom to top). The inset shows the same data in the rescaled axes, $wt^{-\beta}$ vs $lt^{-1/z}$, with the KPZ-class exponents $\beta=1/3$ and $z=3/2$. (b) Overall width $W(t)$ against time $t$. The inset confirms the convergence to the asymptotic power law $W \sim t^\beta$ with $\beta=1/3$. The dashed lines are guides for the eyes indicating the slopes for the KPZ-class exponents.}
  \label{fig:FamilyVicsek}
 \end{center}
\end{figure}

We first check the scaling exponents $\alpha$ and $\beta$
 by the usual method based on the Family-Vicsek scaling
 \cite{Family.Vicsek-JPA1985},
 which describes the spatial and temporal dependence
 of the roughness growth.
To this end, we define the interface width $w(l,t)$
 as the standard deviation of the height $h(x,t)$ measured over length $l$,
 or, more specifically,
 $w(l,t) \equiv \Expct{\sqrt{\expct{\[h(x,t) - \expct{h(x,t)}_l\]^2}_l}}$
 with the ensemble average $\expct{\cdots}$
 and the average $\expct{\cdots}_l$ taken inside a segment of length $l$
 around the position $x$.
The Family-Vicsek scaling then reads
\begin{equation}
 w(l,t) \sim t^\beta F_w(lt^{-1/z}) \sim \cases{l^\alpha & for $l \ll l_*$, \\ t^\beta & for $l \gg l_*$,}  \label{eq:FamilyVicsekWidth}
\end{equation}
 with $z \equiv \alpha/\beta$, a scaling function $F_w$
 and a crossover length scale $l_* \sim t^{1/z}$.
This relation is clearly confirmed in our numerical data,
 together with the KPZ-class exponent values
 $\alpha = 1/2$, $\beta = 1/3$ and $z = 3/2$
 (figure \ref{fig:FamilyVicsek}).
The width $w(l,t)$ at fixed times grows as $w \sim l^\alpha$
 with $\alpha = 1/2$ for short length scales $l$,
 while it saturates at the value that increases with time
 [figure \ref{fig:FamilyVicsek}(a)].
This width at the largest length scale
 can be accurately measured by the overall width
 $W(t) \equiv \sqrt{\expct{[h(x,t)-\expct{h}]^2}}$,
 which indeed shows a clear power law
 $W \sim t^\beta$ with $\beta=1/3$
 [figure \ref{fig:FamilyVicsek}(b)].
Note here that the overall width $W(t)$ should be measured
 with the ensemble average $\expct{\cdots}$
 and not the spatial average taken for each interface,
 which turned out to bias the apparent value of $\beta$
 as detailed in \cite{Takeuchi.Sano-a2012}.
The agreement with the KPZ-class exponents is also checked
 by data collapse: the data for the width $w(l,t)$ at different times
 collapse reasonably well onto a single curve $F_w$
 when $wt^{-\beta}$ is plotted against $lt^{-1/z}$
 [insets of figure \ref{fig:FamilyVicsek}(a)].

The realization of the KPZ-class growth exponent $\beta = 1/3$ implies that
 the local height $h$ of the interfaces can be described
 with a deterministic linear growth term and a stochastic $t^{1/3}$ term,
 as follows:
\begin{equation}
 h \simeq v_\infty t + (\Gamma t)^{1/3} \chi,  \label{eq:Height}
\end{equation}
 where $v_\infty$ and $\Gamma$ denote two constant parameters
 and $\chi$ a random variable that captures
 the fluctuations of the growing interfaces.
Note that we focus for the moment
 on asymptotic one-point statistics of the interface fluctuations
 and hence do not consider the dependence of $\chi$ on $x$ and $t$.

\begin{figure}[t]
 \begin{center}
  \includegraphics[clip]{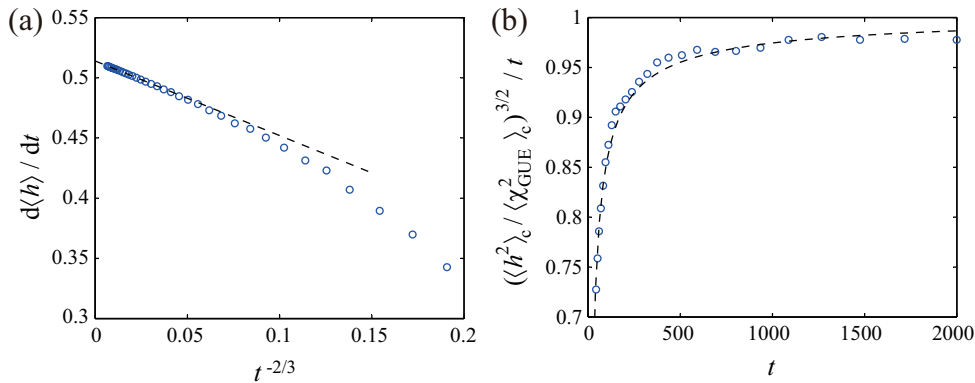}
  \caption{Parameter estimation. (a) Estimation of the linear growth rate $v_\infty$. The time derivative of the mean height, $\rd\expct{h}/\rd t$, is plotted against $t^{-2/3}$. The $y$-intercept of the linear regression (dashed line) provides the estimate of $v_\infty$. (b) Estimation of the amplitude $\Gamma$ for the $t^{1/3}$-fluctuations. The time-series of the second-order cumulant $\cum{h^2}$ is displayed in such a way that the value of $\Gamma$ is found in the limit $t\to\infty$. It is obtained by fitting $at^{-2/3} + \Gamma$ (dashed line) to the data for $t \geq 10$ in order to take into account the finite-time effect (see text).}
  \label{fig:Estimation}
 \end{center}
\end{figure}

First we estimate the values of the two parameters $v_\infty$ and $\Gamma$,
 in the same manner as for the experiment on the liquid-crystal turbulence
 \cite{Takeuchi.Sano-PRL2010,Takeuchi.etal-SR2011,Takeuchi.Sano-a2012}.
The linear growth rate $v_\infty$ is obtained
 as the asymptotic growth speed of the mean height.
More specifically, since equation \eref{eq:Height} indicates
 $\rd\expct{h}/\rd t \simeq v_\infty + c_v t^{-2/3}$ with a constant $c_v$,
 linear regression of the data for $\rd\expct{h}/\rd t$ against $t^{-2/3}$
 in the asymptotic regime, here $t \geq 250$, provides a precise estimate
 of $v_\infty$ at $v_\infty = 0.5139(2)$ [figure \ref{fig:Estimation}(a)].
Concerning the amplitude $\Gamma$ for the $t^{1/3}$-fluctuations,
 we use the relation to the second-order cumulant:
 $\cum{h^2} \equiv W(t)^2 \simeq (\Gamma t)^{2/3} \cum{\chi^2}$.
Setting the arbitrary variance of $\chi$
 to that of the compared distribution function,
 namely the variance of the GUE Tracy-Widom distribution
 denoted by $\cum{\chi_{\rm GUE}^2}$,
 we plot in figure \ref{fig:Estimation}(b)
 $(\cum{h^2}/\cum{\chi_{\rm GUE}^2})^{3/2}/t$
 and read its time asymptotic value.
It shows a considerable finite-time effect,
 which is also visible in the overall width $W(t)$
 [figure \ref{fig:FamilyVicsek}(b)].
Following the method used for the flat interfaces
 in the liquid-crystal experiment \cite{Takeuchi.Sano-a2012},
 we first fit a power law $a_1t^{-\delta} + a_2$
 to the data for $t \geq 10$ in figure \ref{fig:Estimation}(b)
 and obtain $\delta = 0.70(3)$.
Since it is closest to $2/3$ among the multiples of $\beta = 1/3$,
 we assume $\delta = 2/3$ and fit again the data with a weight proportional
 to $(y-a_2)^{-2}$, where $y$ is the ordinate of the data
 as shown in figure \ref{fig:Estimation}(b).
This finally provides our estimate at $\Gamma = 1.01(2)$.
Note that the parameter values obtained here are connected
 to the three parameters in the KPZ equation \eref{eq:KPZEqDef}
 by $v_\infty = \lambda$ \cite{Krug.etal-PRA1992,Takeuchi.Sano-a2012}
 and $\Gamma = A^2|\lambda|/2$ with $A \equiv D/2\nu$
 \cite{Kriecherbauer.Krug-JPA2010}.

\begin{figure}[t]
 \begin{center}
  \includegraphics[clip]{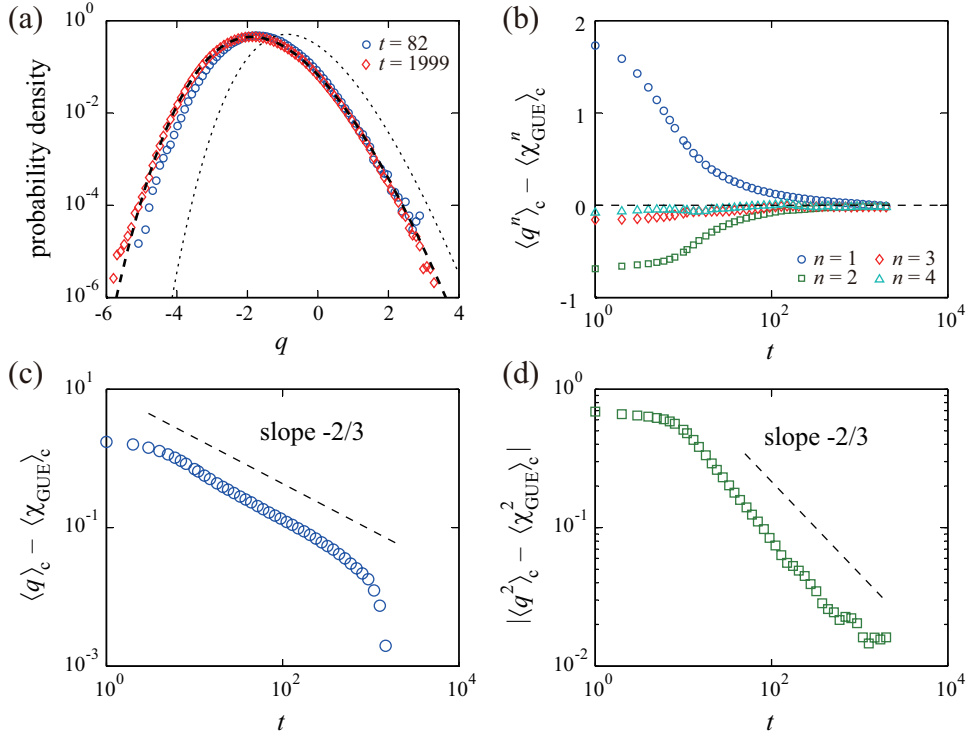}
  \caption{One-point distribution and finite-time corrections. (a) Histogram of the rescaled local height $q \equiv (h - v_\infty t)/(\Gamma t)^{1/3}$ for $t=82$ (blue circles) and $t=1999$ (red diamonds). The dashed line indicates the GUE Tracy-Widom distribution expected for the curved interfaces, whereas the dotted line shows the appropriately rescaled GOE Tracy-Widom distribution for the flat interfaces. (b) Finite-time corrections in the $n$th-order cumulants. Shown are the differences between the cumulants of the measured rescaled height $q$ and the theoretical values for the GUE Tracy-Widom distribution. (c,d) Corrections in the first- and second-order cumulants [same data as (b)] shown in the double logarithmic scales. The dashed lines are guides for the eyes, both indicating the slope $-2/3$.}
  \label{fig:Dist}
 \end{center}
\end{figure}

Using the measured parameter values,
 we rescale the height $h(x,t)$
 in such a way to extract the random variable $\chi$:
\begin{equation}
 q \equiv \frac{h - v_\infty t}{(\Gamma t)^{1/3}} \simeq \chi.  \label{eq:RescaledHeightDef}
\end{equation}
Figure \ref{fig:Dist}(a) displays the histograms
 of the rescaled height $q$ at different times (symbols)
 and compare them with the GUE and GOE Tracy-Widom distributions
 (dashed and dotted lines, respectively).
The data at the later time clearly indicate the GUE Tracy-Widom distribution,
 in agreement with the recent analytical results
 for the curved interfaces in the solvable models
 \cite{Kriecherbauer.Krug-JPA2010,Sasamoto.Spohn-JSM2010,Corwin-RMTA2012}.
This is more quantitatively checked by plotting
 the difference in the $n$th-order cumulant,
 $\cum{q^n} - \cum{\chi_{\rm GUE}^n}$, in figure \ref{fig:Dist}(b).
We thereby confirm that
 the circular interfaces of our off-lattice Eden model
 indeed belong to the KPZ universality class
 at the level of the one-point distribution function.
Moreover, 
 the differences in the $n$th-order cumulants
 also measure the finite-time corrections toward this asymptotic distribution,
 which turn out to decay
 as $|\cum{q^n} - \cum{\chi_{\rm GUE}^n}| \sim t^{-2/3}$ for $n=1$ and $2$
 [figure \ref{fig:Dist}(c,d)].
The power-law decay $t^{-2/3}$ for the second-order cumulant
 has also been reported in all the previously studied systems
 \cite{Ferrari.Frings-JSP2011,Baik.Jenkins-a2011,Sasamoto.Spohn-PRL2010,Sasamoto.Spohn-NPB2010,Takeuchi.Sano-a2012}.
In contrast, noteworthy is the decay $t^{-2/3}$
 found for the first-order cumulant, or the mean,
 because the typical leading correction
 shown by the previous studies is $t^{-1/3}$,
 both for the solvable models
 \cite{Ferrari.Frings-JSP2011,Baik.Jenkins-a2011,Sasamoto.Spohn-PRL2010,Sasamoto.Spohn-NPB2010}%
\footnote{
Ferrari and Frings \cite{Ferrari.Frings-JSP2011} argued
 that the leading correction term for the mean should be in general
 $\mathcal{O}(t^{-1/3})$ for discrete models,
 with mathematical demonstrations for the polynuclear growth (PNG) model
 and the totally and partially asymmetric simple exclusion processes
 (TASEP and PASEP).
They, however, also showed a couple of exceptions where this term vanishes,
 namely the flat PNG interface
 and the PASEP when the hopping rate is tuned to the critical value
 \cite{Ferrari.Frings-JSP2011}.
}
 and for the liquid-crystal experiment
 \cite{Takeuchi.Sano-PRL2010,Takeuchi.etal-SR2011,Takeuchi.Sano-a2012}.
On the one hand, one cannot exclude the possibility
 that the correction in our case shows a crossover
 from $t^{-2/3}$ to $t^{-1/3}$ at larger times,
 as reported in simulations of flat interfaces in the discretized KPZ equation
 \cite{Oliveira.etal-PRE2012}.
On the other hand, our result may indicate that
 our Eden model is endowed with some kind of symmetry,
 for which the leading correction term $t^{-1/3}$ for the mean vanishes.
Note also that Alves \textit{et al.} \cite{Alves.etal-EL2011} numerically found
 the GUE Tracy-Widom distribution in their off-lattice Eden model,
 as well as in two on-lattice isotropic Eden models,
 and reported finite-time corrections for the mean proportional to $t^{-1/3}$.
This difference should arise
 from the different definitions of the off-lattice Eden model
 as described in the previous section.

\begin{figure}[t]
 \begin{center}
  \includegraphics[clip]{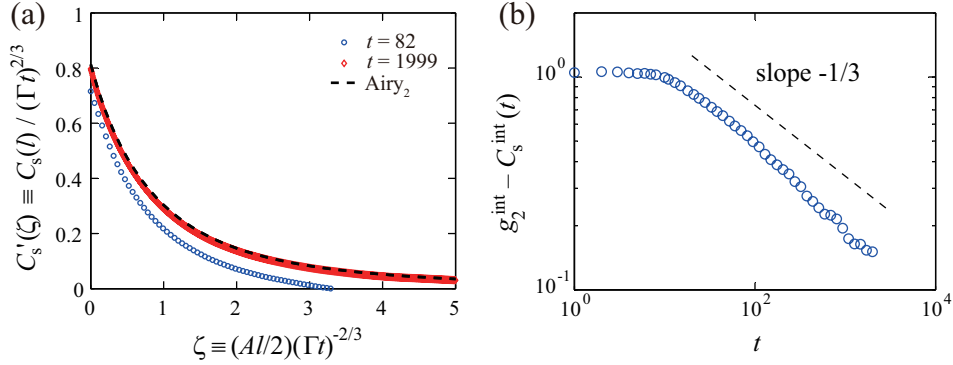}
  \caption{Spatial correlation function $C_{\rm s}(l;t)$. (a) Rescaled correlation function $C'_{\rm s}(\zeta;t) \equiv C_{\rm s}(l;t)/(\Gamma t)^{2/3}$ against rescaled length $\zeta \equiv (Al/2)(\Gamma t)^{-2/3}$. The symbols show the results for the Eden model at $t = 82$ (blue circles) and $t= 1999$ (red diamonds), while the dashed line indicates the Airy$_2$ covariance $g_2(\zeta)$. (b) Finite-time correction in the integral of the rescaled correlation function $C_{\rm s}^{\rm int}(t) \equiv \int_0^\infty C'_{\rm s}(\zeta;t)\rd\zeta$. Displayed is the difference from the value for the Airy$_2$ covariance $g_2^{\rm int} \equiv \int_0^\infty g_2(\zeta)\rd\zeta$ as a function of time. The dashed line is a guide for the eyes showing the slope $-1/3$.}
  \label{fig:SpaceCorr}
 \end{center}
\end{figure}

The spatial correlation function
\begin{equation}
C_{\rm s}(l;t) \equiv \expct{h(x+l,t)h(x,t)} - \expct{h(x+l,t)}\expct{h(x,t)}.  \label{eq:SpatialCorrFuncDef}
\end{equation}
 is also a quantity of central interest in the recent analytical studies
 \cite{Kriecherbauer.Krug-JPA2010,Sasamoto.Spohn-JSM2010,Corwin-RMTA2012}.
For the solved cases of the curved interfaces, it has been shown to coincide
 with the covariance of the stochastic process called the Airy$_2$ process
 $\mathcal{A}_2(t')$ \cite{Prahofer.Spohn-JSP2002},
 which was noticed to be equivalent to the dynamics of the largest eigenvalue
 in Dyson's Brownian motion for GUE random matrices \cite{Johansson-CMP2003}.
The prediction reads
\begin{equation}
 C_{\rm s}(l;t) \simeq (\Gamma t)^{2/3} g_2\( \frac{Al}{2}(\Gamma t)^{-2/3} \)  \label{eq:AiryCorrelation}
\end{equation}
 with $g_2(\zeta) \equiv \expct{\mathcal{A}_2(t'+\zeta)\mathcal{A}_2(t')}$,
 which was indeed substantiated in the liquid-crystal experiment
 \cite{Takeuchi.Sano-PRL2010,Takeuchi.Sano-a2012}.
Our numerical data also confirm this in the time asymptotic limit,
 as shown in figure \ref{fig:SpaceCorr}(a)
 which displays the rescaled correlation function
 $C'_{\rm s}(\zeta;t) \equiv C_{\rm s}(l;t)/(\Gamma t)^{2/3}$ against
 $\zeta \equiv (Al/2)(\Gamma t)^{-2/3}$.
The finite-time correction is also quantified by measuring the integral
 $C_{\rm s}^{\rm int}(t) \equiv \int_0^\infty C'_{\rm s}(\zeta;t)\rd\zeta$
 as a function of time.
This indeed approaches the value for the Airy$_2$ covariance
 $g_2^{\rm int} \equiv \int_0^\infty g_2(\zeta)\rd\zeta$
 by a power law $g_2 - C_{\rm s}^{\rm int}(t) \sim t^{-1/3}$
 [figure \ref{fig:SpaceCorr}(b)], which was also identified
 in the liquid-crystal experiment
 \cite{Takeuchi.Sano-PRL2010,Takeuchi.Sano-a2012}.

From now on we study statistical quantities that remain
 out of reach of rigorous theoretical treatment so far.
Among them particularly important are those characterizing
 the correlation in time, especially the temporal correlation function
\begin{equation}
C_{\rm t}(t,t_0) \equiv \expct{h(x,t)h(x,t_0)} - \expct{h(x,t)}\expct{h(x,t_0)}.  \label{eq:TemporalCorrFuncDef}
\end{equation}
The temporal correlation should be measured along the directions in which
 fluctuations propagate in space-time, called the characteristic lines
 \cite{Ferrari-JSM2008,Corwin.etal-AIHPBPS2012}.
In our case these are simply the radial directions of the circular growth
 and represented by the fixed $x$ in equation \eref{eq:TemporalCorrFuncDef}.

For the circular case, Singha \cite{Singha-JSM2005} performed
 an approximative theoretical calculation
 based on the assumption of effectively linear evolution
 of the height fluctuations,
 and derived
\begin{eqnarray}
 \frac{C_{\rm t}(t,t_0)}{C_{\rm t}(t_0,t_0)} \approx F_{\rm Singha}(t/t_0; b), \label{eq:Singha1} \\
 F_{\rm Singha}(x; b) \equiv \frac{\e^{b(1-1/\sqrt{x})} \Gamma(2/3, b(1-1/\sqrt{x}))}{\Gamma(2/3)}, \label{eq:FSinghaDef}
\end{eqnarray}
 with a single unknown parameter $b$,
 the upper incomplete Gamma function
 $\Gamma(s,x) \equiv \int_x^\infty y^{s-1} \e^{-y} \rd y$
 and the Gamma function $\Gamma(s) = \Gamma(s,0)$.
For the liquid-crystal experiment \cite{Takeuchi.Sano-a2012},
 this functional form works only asymptotically.
More specifically, it fits the experimental data for $t>t_0$
 on condition that
 the right hand side of equation \eref{eq:Singha1} is multiplied
 by a time-dependent coefficient $c(t_0)$,
 which turned out to satisfy $c(t_0) \to 1$ with $t_0 \to \infty$
 and is attributed, somewhat speculatively,
 to microscopic dynamics of the liquid-crystal convection.
On the other hand, Singha himself checked his prediction
 by simulations of the conventional on-lattice Eden model
 \cite{Singha-JSM2005}.
He then found good agreement if an independent value of $b$ is chosen
 for each growth direction,
 without however studying the dependence on $t_0$.

\begin{figure}[t]
 \begin{center}
  \includegraphics[clip]{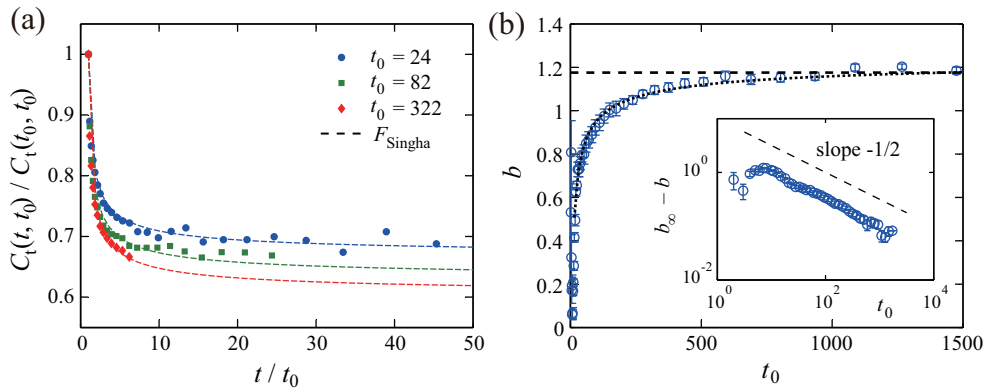}
  \caption{Temporal correlation function $C_{\rm t}(t,t_0)$. (a) $C_{\rm t}(t,t_0) / C_{\rm t}(t_0,t_0)$ against $t/t_0$ with different $t_0$ (as shown in the legend). The dashed lines show the results of fitting by Singha's function \eref{eq:FSinghaDef}. Notice that the correlation function remains far above zero. (b) The values of the fitting parameter $b$ in equation \eref{eq:Singha1} as a function of $t_0$. The dashed line indicates the mean value of $b$ estimated for $t \geq 500$, while the dotted curve shows the result of a power-law fit $b(t_0) = at_0^{-\delta} + b_\infty$. The inset displays the quality of this power-law fit in the axes $b_\infty - b$ vs $t_0$, with a guide for the eyes indicating the slope for $\delta=1/2$.}
  \label{fig:TimeCorr}
 \end{center}
\end{figure}%

\begin{figure}[t]
 \begin{center}
  \includegraphics[clip]{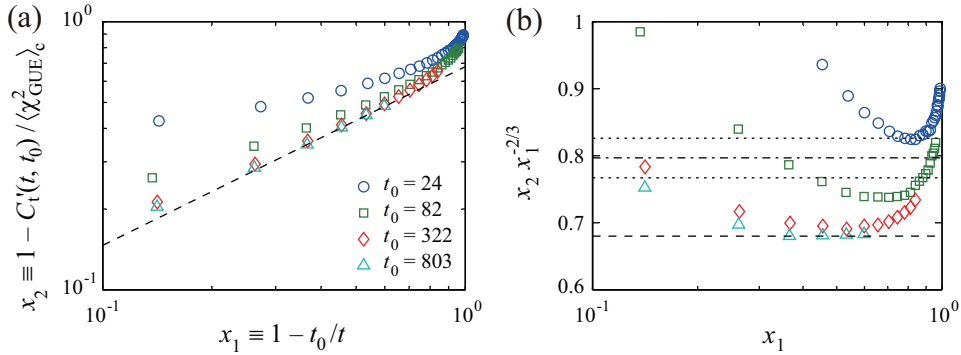}
  \caption{Short-time behavior of the temporal correlation function $C_{\rm t}(t,t_0)$. (a) $x_2 \equiv 1-C'_{\rm t}(t,t_0) / \cum{\chi_{\rm GUE}^2}$ against $x_1 \equiv 1-t_0/t$ with $C'_{\rm t}(t,t_0) \equiv C_{\rm t}(t,t_0) / (\Gamma^2 t_0t)^{1/3}$ for different reference times $t_0$. (b) Same data in the different axes $x_2x_1^{-2/3}$ vs $x_1$. The dashed lines in both panels show the right hand side of equation \eref{eq:TemporalCorrFuncShort} with $R/2 = 0.68$. The dashed-dotted and dotted lines in the panel (b) indicate the value and the confidence interval, respectively, of $3b^{2/3}/2^{5/3}\Gamma(2/3) = 0.80(3)$.}
  \label{fig:TimeCorrShort}
 \end{center}
\end{figure}

The temporal correlation function is measured in our off-lattice Eden model
 and shown in figure \ref{fig:TimeCorr}(a), together with the results
 of fitting by Singha's function \eref{eq:FSinghaDef}.
As shown in the figure, Singha's function fits our data reasonably well
 within the whole time window, or $c(t_0) = 1$ for all $t_0$,
 in agreement with his simulations for the on-lattice Eden model.
We however find that the value of the parameter $b$
 obtained by the best fitting of the data increases with $t_0$
 [figure \ref{fig:TimeCorr}(b)].
Assuming the convergence of $b$ at later times such as $t \geq 500$,
 we estimate the asymptotic value of $b$ at $b = 1.18(3)$.
On the other hand, attempt of a power-law fit in the form
 $b(t_0) = at_0^{-\delta} + b_\infty$ yields a reasonable result
 with $b_\infty = 1.27(3)$
 [dotted curve and inset of figure \ref{fig:TimeCorr}(b)].
Given large uncertainties expected in the nonlinear fit of $F_{\rm Singha}$,
 especially for later times $t_0$,
 here we do not single out either possibility
 but instead provide a single final estimate $b = 1.22(8)$
 that covers both confidence intervals.
In any case, our results indicate that Singha's function \eref{eq:FSinghaDef}
 indeed explains the global form of the temporal correlation function
 in our Eden model.
This implies in particular that positive correlation would remain forever,
 i.e., $\displaystyle{\lim_{t\to\infty} C_{\rm t}(t,t_0) \sim t_0^{2/3} > 0}$,
 which was also inferred for the circular interfaces
 in the liquid-crystal experiment \cite{Takeuchi.Sano-a2012}.
Note that, for the flat interfaces, the temporal correlation was shown to decay
 to zero as $C_{\rm t}(t,t_0) \sim t^{-2/3}$ for fixed $t_0$,
 both numerically \cite{Kallabis.Krug-EL1999}
 and experimentally \cite{Takeuchi.Sano-a2012}.
Concerning the other limit $t \to t_0$,
 equations \eref{eq:Singha1} and \eref{eq:FSinghaDef}
 with $C_{\rm t}(t_0,t_0) = W(t_0)^2 \simeq (\Gamma t_0)^{2/3}\cum{\chi_{\rm GUE}^2}$
 yield
\begin{equation}
 C_{\rm t}(t,t_0) \simeq (\Gamma^2 t_0 t)^{1/3} \cum{\chi^2} \[ 1-\frac{R}{2}\(1-\frac{t_0}{t}\)^{2/3}\] \quad (t-t_0 \ll t_0)  \label{eq:TemporalCorrFuncShort}
\end{equation}
 with $R/2 \approx 3b^{2/3}/2^{5/3}\Gamma(2/3)$ \cite{Singha-JSM2005}.
Our data indeed confirm this functional form (figure \ref{fig:TimeCorrShort}),
 except that they indicate $R/2 \approx 0.68$
 rather than $R/2 \approx 3b^{2/3}/2^{5/3}\Gamma(2/3) = 0.80(3)$.
This implies that Singha's function \eref{eq:FSinghaDef}
 is not quantitatively precise
 in the short-time regime $t-t_0 \ll t_0$ of the temporal correlation function.

In relation to the liquid-crystal experiment,
 our data for the Eden model do not require the extra multiplier $c(t_0)$
 in equation \eref{eq:Singha1},
 which was necessary for the experimental data
 \cite{Takeuchi.Sano-a2012}.
This supports the author's speculation \cite{Takeuchi.Sano-a2012} that
 the need for the coefficient $c(t_0)$ for the experiment
 results from microscopic dynamics of the liquid-crystal convection
 decoupled from the interface growth,
 which is obviously missing in the Eden model.
With this difference in mind, the liquid-crystal experiment
 gave $b \approx 0.8(1)$ and $R/2 \approx 0.6$, to be compared
 with $b = 1.22(8)$ and $R/2 \approx 0.68$ in our Eden model.
This suggests that the parameter $b$
 is not a universal quantity of the KPZ class.

\begin{figure}[t]
 \begin{center}
  \includegraphics[clip]{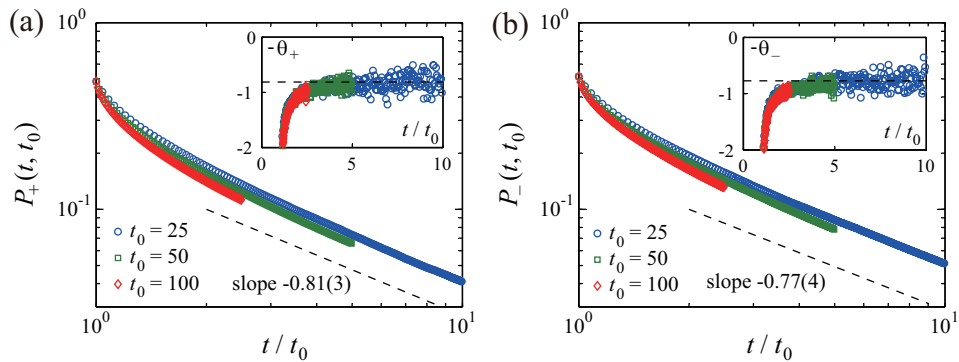}
  \caption{Temporal persistence probability $P_\pm(t,t_0)$ for the positive (a) and negative (b) fluctuations, with different $t_0$ as shown in the legends. The insets show the running exponents $-\theta_\pm(t,t_0) \equiv \rd[\log P_\pm(t,t_0)]/\rd[\log(t/t_0)]$. The dashed lines show the estimated values of the persistence exponents, $\theta_+ = 0.81(3)$ and $\theta_- = 0.77(4)$.}
  \label{fig:TimePersistence}
 \end{center}
\end{figure}

Next we turn our attention to another aspect of the temporal correlation,
 namely the first-passage property
 characterized by the temporal persistence probability $P_\pm(t,t_0)$.
For the fluctuating interfaces, $P_\pm(t,t_0)$ is defined
 as the joint probability that the interface fluctuation
 $\delta h(x,t) \equiv h(x,t) - \expct{h}$
 at a fixed location is positive (negative) at time $t_0$
 and maintains the same sign until time $t$
 \cite{Kallabis.Krug-EL1999,Singha-JSM2005}.
Here we distinguish the positive and negative fluctuations
 denoted by $+$ and $-$, respectively,
 because Kallabis and Krug numerically found power-law decay
 $P_\pm(t,t_0) \sim t^{-\theta_\pm}$ with different exponent values,
 $\theta_+ < \theta_-$ if $\lambda > 0$ and opposite in the other case,
 in the case of flat interfaces \cite{Kallabis.Krug-EL1999}.
This asymmetry was also confirmed in the liquid-crystal experiment
 for the flat interfaces, while no asymmetry was identified
 for the circular case \cite{Takeuchi.Sano-a2012}.
Our numerically produced circular interfaces also support
 this claim, as summarized in figure \ref{fig:TimePersistence}.
In the main panels, we find power-law decay
\begin{equation}
 P_\pm(t,t_0) \sim (t/t_0)^{-\theta_\pm}  \label{eq:PersistenceProb}
\end{equation}
 for sufficiently large $t/t_0$.
This is further confirmed in the insets,
 where the running exponents
 $-\theta_\pm(t,t_0) \equiv \rd[\log P_\pm(t,t_0)]/\rd[\log(t/t_0)]$
 with different $t_0$ overlap reasonably well when plotted against $t/t_0$
 and converge to constants for large abscissae.
Fitting the data within this plateau regime
 (not only for the three $t_0$ in figure \ref{fig:TimePersistence}),
 we obtain $\theta_+ = 0.81(3)$ and $\theta_- = 0.77(4)$.
This is in good agreement with the experimentally found values of
 $\theta_+ = 0.81(2)$ and $\theta_- = 0.80(2)$ for the circular interfaces
 and significantly different from those for the flat interfaces,
 $\theta_+ = 1.35(5)$ and $\theta_- = 1.85(10)$ \cite{Takeuchi.Sano-a2012}.
Our result confirms in particular the absence of the asymmetry
 between the positive and negative fluctuations for the circular interfaces,
 i.e., $\theta_+ = \theta_-$ in our precision.
The asymmetry $\theta_+ \neq \theta_-$ present for the flat interfaces
 has been attributed to the nonlinear term
 of the KPZ equation \eref{eq:KPZEqDef} \cite{Kallabis.Krug-EL1999};
 understanding how this effect is cancelled for the circular interfaces
 is an interesting problem left for future studies.

\begin{figure}[t]
 \begin{center}
  \includegraphics[clip]{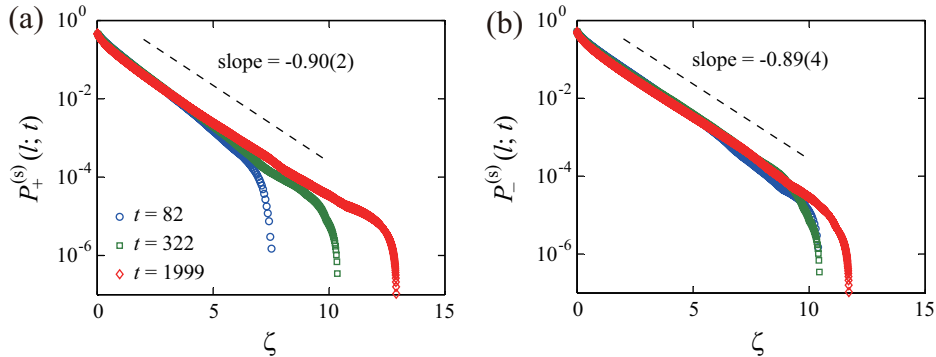}
  \caption{Spatial persistence probability $P_\pm^{\rm (s)}(l;t)$ for the positive (a) and negative (b) fluctuations, with different $t$ as shown in the legends. The length $l$ is shown in the rescaled unit $\zeta \equiv (Al/2)(\Gamma t)^{-2/3}$. The dashed lines are guides for the eyes showing the estimated values of $\kappa^{\rm (s)}_+$ and $\kappa^{\rm (s)}_-$.}
  \label{fig:SpacePersistence}
 \end{center}
\end{figure}

The notion of the persistence can also be considered in space
 \cite{Majumdar.Bray-PRL2001,Constantin.etal-PRE2004}.
It is then characterized by the spatial persistence probability
 $P_\pm^{\rm (s)}(l;t)$,
 which is the probability that a fluctuation remains positive or negative
 over length $l$ in a spatial profile of the interfaces at time $t$.
Although a few analytical and numerical studies have shown
 nontrivial character of the spatial persistence for the stationary interfaces
 \cite{Majumdar.Bray-PRL2001,Constantin.etal-PRE2004},
 for the growing interfaces it has been studied only experimentally so far,
 in the liquid-crystal experiment \cite{Takeuchi.Sano-a2012}
 and in an earlier experiment on paper smoldering \cite{Merikoski.etal-PRL2003}
 which also exhibited the KPZ scaling exponents \cite{Maunuksela.etal-PRL1997}.
Interestingly, the two experiments showed different results:
 exponential decay for the liquid-crystal turbulence and power-law decay
 $P_\pm^{\rm (s)}(l;t) \sim l^{-1/2}$ for the paper smoldering.
The spatial persistence is therefore measured in our Eden model
 and yields the results shown in figure \ref{fig:SpacePersistence}.
We then find clear exponential decay
 for both positive and negative fluctuations,
\begin{equation}
 P_\pm^{\rm (s)}(l;t) \sim \e^{-\kappa^{\rm (s)}_\pm \zeta},  \label{eq:SpatialPersistence}
\end{equation}
 with values of $\kappa^{\rm (s)}_\pm$ independent of $t$
 when plotted against 
 the dimensionless length scale $\zeta \equiv (Al/2)(\Gamma t)^{-2/3}$.
Measuring them
 at the latest two times at which the interface profiles were recorded,
 namely $t = 1717$ and $1999$, we obtain estimates of
 $\kappa^{\rm (s)}_+ = 0.90(2)$ and $\kappa^{\rm (s)}_- = 0.89(4)$
 for our Eden model.
They are in excellent agreement with the values numerically obtained
 for the temporal persistence of GUE Dyson's Brownian motion,
 $\kappa^{\rm (s)}_+ = 0.90(8)$ and $\kappa^{\rm (s)}_- = 0.90(6)$
 \cite{Takeuchi.Sano-a2012},
 which is expected to be equivalent to the spatial profile
 of the curved KPZ-class interfaces.
This is therefore evidence for universal spatial persistence
 in the growth regime of the KPZ class,
 which in fact concerns infinite-point correlation
 in the spatial profile.
It is also noteworthy that the positive and negative fluctuations
 show no asymmetry in this spatial persistence either.
Similar values were also reported in the liquid-crystal experiment,
 $\kappa^{\rm (s)}_+ = 1.07(8)$ and $\kappa^{\rm (s)}_- = 0.87(6)$
 \cite{Takeuchi.Sano-a2012}, albeit with a small difference
 in the former which is probably due to finite-time effect
 but should be clarified by further study.
In passing, for the flat interfaces, the liquid-crystal experiment showed
 $\kappa^{\rm (s)}_+ = 1.9(3)$ and $\kappa^{\rm (s)}_- = 2.0(3)$
 \cite{Takeuchi.Sano-a2012},
 which are significantly different from the values of the circular case.

\begin{figure}[t]
 \begin{center}
  \includegraphics[clip]{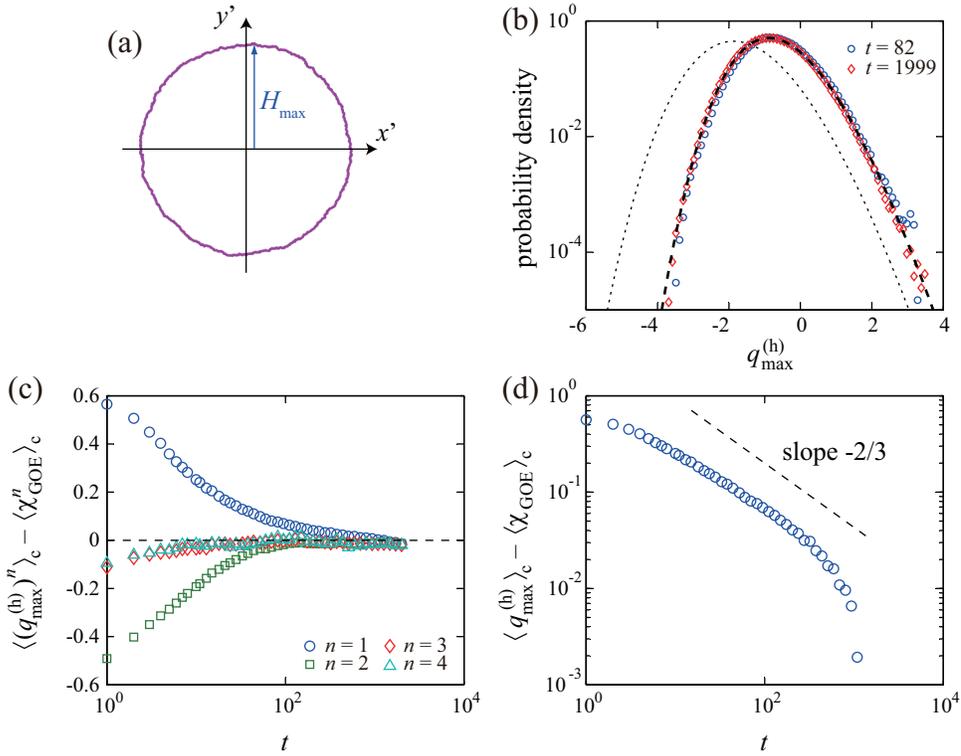}
  \caption{Maximal height distribution. (a) Definition of the maximal height $H_{\rm max}$. The origin of the frame $(x',y')$ is fixed at the position of the initial seed of the cluster. Shown in purple are the active particles of a cluster at $t=1999$, which is also displayed in figure \ref{fig:Model}(b). (b) Histogram of the rescaled maximal height $q^{\rm (h)}_{\rm max} \equiv (H_{\rm max} - v_\infty t)/(\Gamma t)^{1/3}$ for $t=82$ (blue circles) and $t=1999$ (red diamonds). The dashed and dotted lines indicate the GOE and GUE Tracy-Widom distributions, respectively. Note that the conventional definition of the GOE TW random variable is multiplied by $2^{-2/3}$, in view of the theoretical results for the solvable models \cite{Johansson-CMP2003,Forrester.etal-NPB2011,Corwin.etal-a2011,Liechty-a2011,Schehr-a2012}. (c) Finite-time corrections in the $n$th-order cumulants. Shown are the differences between the cumulants of the measured $q^{\rm (h)}_{\rm max}$ and the theoretical values for the GOE Tracy-Widom distribution. (d) Correction in the first-order cumulant [same data as (c)] shown in the double logarithmic scales. The dashed line is a guide for the eyes indicating the slope $-2/3$.}
  \label{fig:MaxHeight}
 \end{center}
\end{figure}

We finally study extreme-value statistics for the interface fluctuations
 with our numerical data.
Specifically, we focus on the maximal height $H_{\rm max}$
 measured with respect to a fictitious substrate that extends from the origin
 [figure \ref{fig:MaxHeight}(a)],
 which is known to obey, asymptotically,
 the GOE Tracy-Widom distribution by studies of solvable models
 \cite{Johansson-CMP2003,Forrester.etal-NPB2011,Corwin.etal-a2011,Liechty-a2011,Schehr-a2012}%
\footnote{
More precisely, the conventional definition
 for the GOE Tracy-Widom random variable should be multiplied by $2^{-2/3}$
 to agree with the naturally rescaled maximal height
 \cite{Johansson-CMP2003,Forrester.etal-NPB2011,Corwin.etal-a2011,Liechty-a2011,Schehr-a2012}.
Our definition of the variable $\chi_{\rm GOE}$
 includes this factor $2^{-2/3}$.
}.
This is the same distribution as for the one-point distribution
 of the \textit{flat} interfaces and
 differs from the GUE Tracy-Widom distribution
 for that of the circular interfaces.
For our Eden model, we measure this maximal height directly
 by $H_{\rm max} = \max y'_i$,
 where $i$ denotes the index of the active particles
 and $(x',y')$ the coordinates defined with respect to the substrate
 [figure \ref{fig:MaxHeight}(a)].
Statistics of $H_{\rm max}$ is improved by rotating the substrate,
 or the frame $(x',y')$, arbitrarily around the origin.
The obtained histograms are shown in figure \ref{fig:MaxHeight}(b)
 for the rescaled maximal height
 $q_{\rm max}^{\rm (h)} \equiv (H_{\rm max} - v_\infty t)/(\Gamma t)^{1/3}$,
 which clearly confirm the asymptotic GOE Tracy-Widom distribution
 for this quantity.
Moreover, in figure \ref{fig:MaxHeight}(c,d),
 we measure the finite-time corrections
 in the $n$th-order cumulants $\cum{(q_{\rm max}^{\rm (h)})^n}$
 and find
 $\cum{q_{\rm max}^{\rm (h)}} - \cum{\chi_{\rm GOE}} \sim t^{-2/3}$
 for the mean,
 in contrast with $t^{-1/3}$
 found for the liquid-crystal experiment \cite{Takeuchi.Sano-a2012}.
Since the same set of the different exponents has been found
 for the finite-time correction in the one-point distribution,
 the author believes that the leading correction terms $t^{-1/3}$
 for the mean of both distributions
 vanish for the same reason in our Eden model.

\section{Summary}

In the present paper, we have introduced an off-lattice variant
 of the Eden model and determined the statistical properties
 of its circular interfaces, as a numerical case study
 for the $(1+1)$-dimensional KPZ universality class.
Besides confirming the universal distribution function
 and the spatial correlation function in agreement with
 the analytical results on the solvable models,
 namely the GUE Tracy-Widom distribution and the Airy$_2$ covariance,
 our particular focus has been put on the unsolved statistical properties
 such as the temporal correlation function and the persistence probability
 in space and time.
Good agreement is then found with the recent experiment
 on the growing interfaces in the liquid-crystal turbulence,
 as far as characteristic quantities are compared
 such as the temporal persistence exponent $\theta_\pm$ and
 the exponential decay rate $\kappa_\pm^{\rm (s)}$ for the spatial persistence.
In particular, unlike the flat interfaces, we have demonstrated that
 $\theta_+ = \theta_-$ holds for the circular case,
 supporting one of the conclusions reached in the liquid-crystal experiment.
The temporal correlation function has also been shown to fit
 Singha's approximative functional form
 \eref{eq:Singha1} and \eref{eq:FSinghaDef}
 without the small modification required for the liquid-crystal experiment.
This indicates that the circular interfaces
 have long-lasting temporal correlation,
 i.e., $C_{\rm t}(t,t_0) > 0$ for large $t$,
 presumably even in the limit $t\to\infty$ unlike the flat interfaces.
These agreements with the experimental results for the circular case
 and the sharp contrast to the flat interfaces
 (see table 1 of \cite{Takeuchi.Sano-a2012} for a summary)
 further emphasize the geometry-dependent universality
 of the $(1+1)$-dimensional KPZ class.

In view of these experimental and numerical results
 for the universality beyond the analytically solved quantities,
 it would be important to provide firmer theoretical grounds
 for such unsolved universal statistical properties,
 especially those related to the temporal correlation.
This would call for several complementary approaches: 
 attempts to provide analytical solutions for those quantities
 in solvable models,
 refinement of approximative theoretical evaluation
 performed by Singha \cite{Singha-JSM2005},
 and recent developments in the application of renormalization group techniques
 \cite{Canet.etal-PRL2010,Canet.etal-PRE2011,Corwin.Quastel-a2011}
 are also intriguing.
Combination of such theoretical approaches
 as well as further experimental and numerical case studies
 would surely help understand this prominent out-of-equilibrium universality
 of the KPZ class,
 which governs the general phenomenon of the growing interfaces.

\ack

The author is grateful to T.~Sasamoto
 for regular and fruitful discussions exchanged with the author.
The author acknowledges the theoretical curves of the Tracy-Widom distributions
 provided by M.~Pr\"ahofer and those of the Airy$_2$ covariance
 by F.~Bornemann, the latter of which is evaluated
 with Bornemann's accurate algorithm \cite{Bornemann-MC2010}.
This work is supported in part
 by Grant for Basic Science Research Projects from The Sumitomo Foundation
 and by the JSPS Core-to-Core Program ``International research network
 for non-equilibrium dynamics of soft matter.''

\section*{References}

\bibliographystyle{iopart-num}
\bibliography{eden,otherrefs}

\end{document}